\documentclass[10pt,compsoc]{IEEEtran}

\usepackage[nocompress]{cite}
\usepackage{amsthm,amssymb,amsmath}
\usepackage{graphicx}

\theoremstyle{plain}
\newtheorem{theorem}{Theorem}

\theoremstyle{definition}
\newtheorem{alg}{Algorithm}
\newtheorem{remark}{Remark}
\newtheorem{example}[remark]{Example}

\newcommand{\desc}{\operatorname{descendants}}
\newcommand{\MRCA}{\operatorname{MRCA}}
\newcommand{\tc}{\text{:}} % tc=tight colon, for use in math mode in Newick trees

\begin{document}

\title{Topological metrizations of trees, and new quartet methods of tree inference}
\author{John A. Rhodes
\thanks{J.A. Rhodes is with the Department of Mathematics and Statistics, University of Alaska, Fairbanks; email: j.rhodes@alaska.edu}
}
%\address{Department of Mathematics and Statistics, University of Alaska Fairbanks, 99775}

%\email{j.rhodes@alaska.edu}

%\date{October 10, 2018}

%\keywords{Phylogenetic tree; Quartet; Rooted Triple; Species Tree, Incomplete Lineage Sorting, Multispecies Coalescent}

%\subjclass{92D15}

\maketitle

\begin{abstract}
Topological phylogenetic trees can be assigned edge weights in several natural ways, highlighting different aspects of the tree. Here the rooted triple and quartet metrizations are introduced, and applied to formulate novel methods of inferring large trees from rooted triple and quartet data. These methods lead to new statistically consistent procedures for  inference of a species tree from gene trees under the multispecies coalescent model.
\end{abstract}

\begin{IEEEkeywords}
Phylogeny, Genomics, Evolution.
\end{IEEEkeywords}

%%%%%%%%%%%%%%%%%%%%%%%%%%%%%%%%%%
%%%%%%%%%%%%%%%%%%%%%%%%%%%%%%%%%%
%%%%%%%%%%%%%%%%%%%%%%%%%%%%%%%%%%
%%%%%%%%%%%%%%%%%%%%%%%%%%%%%%%%%%

\section{Introduction}

The inference of a species trees, which shows the  evolutionary relationships between a collection of taxa, from gene trees, which depict the joining of ancestral lineages for genes sampled from individuals of those taxa, is made difficult by the fact that gene tree topologies often differ from each other. One biological explanation for such gene tree incongruity  is
\emph{incomplete lineage sorting} (ILS). In ILS, gene lineages may not share a common ancestor in the most recent ancestral population in which it is possible for them to do so. This allows the lineages to merge with those from more distantly-related species before they do so with closer ones. The formation of gene trees within species trees taking into account ILS is described by the \emph{multispecies coalescent model} (MSC). Though many inference methods have been proposed to recover a species tree from a collection of gene trees under the MSC, both computational requirements and performance in simulation vary enough that no single approach has yet become clearly preferred.

In \cite{Liu2009} and \cite{Liu2011}, Liu and coworkers proposed particularly interesting and fast methods for this inference problem, using as data collections of unrooted and rooted gene trees, respectively. These methods, called STAR and $\text{NJ}_{st}$, proceed by 
first discarding any metric information on the gene trees, and then remetrizing them in a way that reflects only their topological structure. For instance, in the second of these works, the metrization of an unrooted gene tree is to simply make all edges have length 1.  A table of intertaxon distances is then constructed for each gene tree, the mean distance table across the gene trees is computed, and this mean table is used to infer a species tree by  a standard distance approach such as Neighbor Joining. In other words,  average consensus \cite{AvCon,bryant2003classification} is applied to the remetrized gene trees. Although far from intuitively clear, the statistical consistency of these methods has been established \cite{adr2013,adr2018}. Moreover, the strong performance of the methods on simulated data has been shown both in the original works, and by the implementation of $\text{NJ}_{st}$ in the software ASTRID \cite{ASTRID}.

Although not discussed in \cite{Liu2009} or \cite{Liu2011}, but as explained below, the two remetrizations of gene trees these works use  can be  understood by viewing them as related to the notions of clades and splits on trees. That is, the intertaxon distances on a remetrized gene tree provides a numerical summary of the tree built on these specific combinatorial notions. Since clades and splits are only two of the combinatorial  tools useful for describing topological trees, natural questions are 1) what other combinatorial notions lead to metrizations that usefully capture topological information about trees?, and 2) how can these metrizations be used for tree inference? The goal of this note is to investigate these questions for the notions of displayed rooted triples (induced rooted 3-leaf  trees) and quartets (induced unrooted 4-leaf trees).

\medskip

The rooted triple and quartet metrizations developed here  have an important feature: Intertaxon distances can be computed from a list of rooted triples or quartets displayed on a tree --- without knowing the tree itself.  This allows them to be used in new supertree methods, which take a collection of rooted triples or quartets, compute a pairwise distance table from them, and then construct a large tree which fits this distance table. If the input rooted triples or quartets are the full set of those displayed on a large tree, this recovers the tree.
Even if some of the triples or quartets are erroneous or missing, approximations to  the intertaxon distances on the tree are still obtained, and a tree can be quickly inferred by any of the well-known distance-based methods for tree building or selection that are robust to error. We call these methods Quartet Distance Supertree (QDS) and Rooted Triple Distance Supertree (RTDS).

This approach extends to give new statistically consistent methods of species tree inference from samples of topological gene trees drawn from the MSC model, adding to those few already known \cite{DegnanConsensus2009,adr2013,adr2018}. The key additional ingredient for this inference is the fact that under the MSC the most frequent rooted triple or quartet topology across a collection of independent genes reflects that of the species tree. From a collection of gene trees, one can tabulate frequencies of the displayed quartets on four taxa, choosing the most frequent as the inferred species quartet on those taxa. Then QDS can be used for statistically consistent inference of the full species tree. We refer to this method, and its analog using rooted triples, as Quartet Distance Consensus (QDC) and Rooted Triple Distance Consensus (RTDC).

While QDC for species tree inference should certainly not be expected to have the speed of $\text{NJ}_{st}$, due to its need to consider quartets individually, it offers another advantage. Specifically, it remains statistically consistent even when some gene trees have some taxa missing. Although the quartet-based scheme implemented in the software ASTRAL-III should also be less vulnerable to problems with missing taxa on gene trees, QDC may in some circumstances offer advantages over it as well. In particular, a theoretical complexity analysis indicates that while QDC's running time has higher exponent on the number of taxa than does ASTRAL-III's, it has lower exponent on the number of gene trees. Thus for a moderate number of taxa but a large number of gene trees, an efficient implementation may offer superior runtimes.

Although the focus of this paper is on the development of the quartet and rooted triple metrizations, using simulated datasets previously used to evaluate ASTRID and ASTRAL-III we provide some evidence on the performance of QDC in comparison to $\text{NJ}_{st}$ and ASTRAL-III. Determining whether this behavior is typical, however, requires further work. An efficient software implementation and large-scale simulation studies of performance  are necessary for a more complete assessment. Regardless of the ultimate performance of these algorithms, though, the distances on which they are built may find other uses in phylogenetic theory and practice.

\medskip

Basic definitions are given in Section \ref{sec:def}. The clade and split metrizations are presented in detail in Section \ref{sec:sc}. The main theoretical contributions of this note are the two new metrizations associated to rooted triples and quartets developed in Sections \ref{sec:rt} and \ref{sec:q}. The applicability of this theory to 
supertree inference from rooted triple and quartets and to species tree inference from gene trees is then developed in Sections \ref{sec:super} and \ref{sec:st}, where simulation results are also given. Section \ref{sec:conc} concludes with a few general comments.

\section{Notation and terminology}\label{sec:def}

Throughout this work $X$ denotes a finite set of $N$ taxa. Upper case letters $A,B,\dots$ denote subsets of $X$, and lower case letters $a,b,\dots$  elements of $X$.

A \emph{split of $X$} is a bipartition $X=A\sqcup B$ of the taxa into non-empty subsets, and is denoted $A|B =B|A$. A \emph{clade} of $X$ is a non-empty subset $A\subseteq X$ of the taxa. A resolved \emph{rooted triple of $X$} is a subset of three elements of $X$, partitioned into a pair $a,b$ and a singleton $c$, and denoted $ab|c=ba|c$. 
To allow for multifurcations on trees, we will also have need for an unresolved rooted triple $abc$.
A resolved \emph{quartet} of $X$ is a subset of four elements of $X$, partitioned into two pairs $a,b$ and $c,d$, and denoted $ab|cd=ba|cd=\cdots =cd|ab$. An unresolved quartet is $abcd$.

Suppose the taxa $X$ bijectively label the leaves of a rooted  tree $T^r,$ or of an unrooted tree $T$, with the root of degree at least $2$ and all other internal nodes of degree at least $3$.  Then $T^r$ and $T$ are said to be \emph{phylogenetic trees on $X$}.  All edges on a rooted tree $T^r$ are directed away from the root, so, for instance, the root is ancestral to all leaves. Edges on an unrooted tree $T$ are undirected. A phylogenetic tree is \emph{binary} if the minimal degree conditions on the nodes are met, and is otherwise said to be \emph{polytomous}.

A tree $T^r$ \emph{displays the clade} $A$ if the most recent common ancestor (MRCA) on $T^r$ of the taxa in $A$ has as its descendants in $X$ precisely the set $A$. Thus clades displayed on a rooted  tree correspond to its nodes, and if the tree is binary, it displays exactly $2N-1$ clades, including all singleton clades and the clade $X$. We say $T^r$ \emph{displays the rooted triple} $ab|c$ if the MRCA of $a$ and $b$ is a proper descendent of the MRCA  of $a,$ $b,$ and $c$. In the case of a rooted polytomous tree, we say
 the unresolved rooted triple $abc$ is displayed if the MRCAs of the three pairs $a,b$; $a,c$; and $b,c$ coincide.
If $N\ge 3$, a rooted tree on $X$ thus displays $\binom N3$ rooted triples, one for each choice of three taxa. 

Similarly, for an unrooted tree $T$ on taxa $X$, we say $T$ \emph{displays the split} $A|B$ if the bipartition is obtained by removing an edge of $T$ and partitioning $X$ according to connected components of the resulting graph. If $N\ge 2$, a binary tree displays $2N-3$ splits. We say $T$ \emph{displays the quartet} $ab|cd$ if 
on the induced 4-leaf tree relating $a$, $b$, $c$, and $d$ the split $\{a,b\}|\{c,d\}$ is displayed. A tree displays the unresolved quartet $abcd$ if the induced tree relating them is a star tree.
If $N\ge 4$ a tree thus displays $\binom N4$ quartets, one for each subset of four taxa. 

Suppose positive weights are somehow assigned to the edges of $T^r$ or $T$, so the tree is now a \emph{metric tree}. Any such edge weighting scheme $W$ induces a metric $d_W$ on $X$, using the sum of edge weights along paths between pairs of taxa. As is well known, however, a metric $d$ on $X$ need not arise from such a weighting. If $d=d_W$ for some $W$ on $T$ or $T^r$, then we say $d$ is a \emph{tree metric} on $T$ or $T^r$ with weighting $W$.

\section{Split and clade metrizations of trees}\label{sec:sc}

For completeness and perspective on what is to follow, we present two topological metrizations of trees that have been used in other works.

\medskip

Given an unrooted topological tree $T$ on $X$, we may assign  weights $w(e)=1$ to all edges $e$. The resulting tree metric on $X$ is just the usual graph-theoretic distance along $T$.  However, by the correspondence between displayed splits and edges on the tree, the distance between taxa $x$ and $y$ can also be described as  the number of splits $A|B$ displayed on $T$ that \emph{separate} $x,y$, in the sense that $x\in A$, $y\in B$.
For this reason, we denote the weighting scheme with all weights 1 by $Sp$, and say it gives the \emph{split metrization} of $T$. This is essentially the metrization used in \cite{Liu2011} for the $\text{NJ}_{st}$ algorithm for species tree inference, which was renamed U-STAR/NJ  in \cite{adr2018} since the same distance approach generalizes to U-STAR/M for any distance method M of tree construction or selection. The ASTRID software \cite{ASTRID} is an implementation of these U-STAR methods.

\medskip

For a rooted topological tree $T^r$ on $X$, assign numbers to the internal nodes of the tree as follows: To the root assign $N$, to its children that are internal nodes assign $N-1$, to their children that are internal nodes assign $N-2$, and so on, decreasing by 1 for each parent-to-child step. Assign $0$ to all leaves. Then assign edge weights $w(e)$ as the positive difference of the numbers on the endpoints of $e$.  Thus all internal edges are weighted 1, but terminal edges are weighted with possibly different numbers between 2 and $N$.  All leaf-to-root distances are $N$, so the tree is ultrametric. We denote this weighting scheme by $Cl$, say it gives the \emph{clade metrization} of $T^r$, and denote the induced metric on $X$ by $d_{Cl}$.
The name is justified by the observation in \cite{adr2013} that for $x,y\in X$, $$d_{Cl}(x,y)=2(1+N-|C_{x,y}| )$$ where $C_{x,y}$ is the set of clades displayed on $T^r$ that contain both $x$ and $y$. This follows directly by the correspondence between nodes on the path between the tree root and  $\MRCA(x,y)$ to displayed clades containing $x$ and $y$.
This clade metrization was introduced in \cite {Liu2009} for the STAR algorithm for species tree inference.

\begin{remark}
As shown in \cite{adr2013}, there are generalizations of the clade metrization which can be used for consistent inference of species trees from gene trees following the plan of \cite{Liu2009}. These generalizations, which allow for the weight of an edge to depend on the number of edges between it and the root, are not used in this paper.\end{remark}

\begin{remark}
One might propose an alternative clade metrization, $Cl'$ defined by assigning unit weights to all edges in a rooted tree. 
Then the intertaxon distance 
$d_{Cl'}(x,y)$ is the number of clades displayed on the tree that contain one of $x,y$ but not both. However, $d_{Cl'}$ lacks the ultrametricity of $d_{Cl}$. It can also
be obtained by restricting the split distance on the larger unrooted tree
obtained by attaching to the root a single edge leading to an extra taxon.
\end{remark}

\section{Rooted triple metrization of a rooted tree}\label{sec:rt}
With $T^r$ a rooted phylogenetic tree on $X$, we may assign edge weights to $T^r$ as follows:
First number each node of the tree, including leaves, with the number of taxa descended from it. Leaves are numbered 1, as they are considered their own descendants, and the root is numbered $N$, the total number of taxa. Then assign weights $w(e)$ as the positive difference of the numbers on the endpoints of $e$. That is, for any edge $e=(u,v)$ directed away from the root with $u$ the parent of $v$, the edge weight is
$$w_{RT}(e)=|\desc(u)|-|\desc(v)|,$$ the decrease in number of descendants across $e$. We refer to this as the \emph{rooted triple metrization}, a name  justified by Theorem \ref{thm:drt} below, and denote the weighting scheme $RT$. It results in an ultrametric tree, with the root at distance $N-1$ from every leaf, and more generally every internal node $u$ at distance 
$|\desc(u)|-1$ from its leaf descendants.

\begin{example}
With the rooted triple metrization, a rooted caterpillar tree $$(\dots (((a_1,a_2),a_3),a_4), \dots, a_N)$$ will have all internal edges of weight 1. The pendant edges, listed from the cherry toward the root, will have weights $1,1,2,3,4,\dots, N-1$. 
\end{example}

\begin{example}
With the rooted triple metrization, a  balanced tree $$(\dots ((a_1,a_2),(a_3,a_4)),\dots, ((a_{N-3},a_{N-2}),(a_{N-1},a_N))\dots )$$
on $N=2^k$ taxa will have pendant edges of weight 1, and as one moves toward the root, internal edges of weight $2,4,8,\dots, 2^{k-1}$.
\end{example}

\begin{theorem}\label{thm:drt}
Suppose a rooted phylogenetic tree $T^r$ is given the rooted triple metrization. Then the resulting tree metric $d_{RT}$ satisfies, for all $x,y\in X$, $x\ne y$,
$$d_{RT}(x,y)= 2|R_{x,y}|+2,$$
where $R_{x,y}$ is the set of rooted triples displayed on $T$ of the form $xz|y$, $yz|x$, or $xyz$. 
\end{theorem}
More informally, for a binary tree $T$ the distance $d_{RT}(x,y)$ is, up to a simple transformation, the number of rooted triples displayed on $T$ in which $x,y$ are \emph{separated}. This remains true for trees with polytomies as long as unresolved triples are viewed as separating all taxa in them.

\begin{proof}
Let $v=\MRCA(x,y)$, and $k$ be the number of leaf descendants of $v$ (i.e., $k$ is the size of the smallest displayed clade containing $x,y$). Since $d_{RT}(x,y)$ is the sum of edge weights on the path between $x$ and $y$, we find that 
\begin{align*}d_{RT}(x,y)&= d_{RT}(x,v)+d_{RT}(v,y)\\
&=(k-1)+(k-1)=2k-2.
\end{align*}
The number of rooted triples of the forms $xz|y$, $yz|x$, or $xyz$ is the number of taxa $z$ descended from $v$, excluding $x$ and $y$. Thus 
$$|R_{x,y}|=k-2.$$
Eliminating $k$ from these two equations yields the claim.
\end{proof}

\begin{remark}
Combining Theorem \ref{thm:drt}  with the fact that one can determine a rooted ultrametric tree from intertaxon distances on it gives an alternative proof of the well-known result that the collection of rooted triples displayed on a tree determines the rooted tree topology.
\end{remark}

\section{Quartet metrization of a unrooted tree}\label{sec:q}
Let $T$ be an unrooted binary tree on $X$. Each internal edge of $T$ determines a partition of $X$ into 4 non-empty blocks, $X_1,X_2,X_3,X_4$ where the split associated to the edge is $X_1\cup X_2|X_3\cup X_4$, and the splits associated to the 4 adjacent edges all have an $X_i$ as one split set. We refer to this partition as the \emph{quartet partition} associated to an internal edge, and denote it by $X_1,X_2| X_3,X_4$. 
Assign an internal edge $e$ with quartet partition  $X_1,X_2| X_3,X_4$ the weight $$w_Q(e)=|X_1||X_2|+|X_3||X_4|.$$  
For a pendant edge to leaf $x$, the non-leaf endpoint determines a tripartition of the taxa, $\{x\},X_1,X_2$. Assign to such an edge $e$ the weight $$w_Q(e)=|X_1||X_2|.$$

This weighting scheme can be extended to polytomous trees as follows:
Suppose an edge $e=\{u,v\}$ is incident to $m_u$ other edges $e_i$ at $u$, and $m_v$ other edges $\tilde e_j$ at $v$. Note that for a leaf $v$ we allow $m_v=0$.
Now let $X_1,X_2,\dots X_{m_u} ,\tilde X_1,\tilde X_2,\dots, \tilde X_{m_v}$ be the partition of $X$ with $X_i$ (respectively $\tilde X_j$) the set of taxa connected to $e$ by a path through $e_i$ (respectively $\tilde e_j$). Then assign to $e$ the edge weight
$$w_Q(e)=\sum_{1\le i< i'\le m_u} |X_i||X_{i'}| +\sum_{1\le j<j'\le m_v} |\tilde X_j||\tilde X_{j'}|.$$
Interpreting an empty sum as 0, this agrees with the definition above for binary trees.

We refer to this as the \emph{quartet metrization}, due to Theorem \ref{thm:dq} below, and denote the weighting scheme $Q$. Trees with the quartet metrization are usually not ultrametric, as examples show.

\begin{figure}
\includegraphics[height=2.25in]{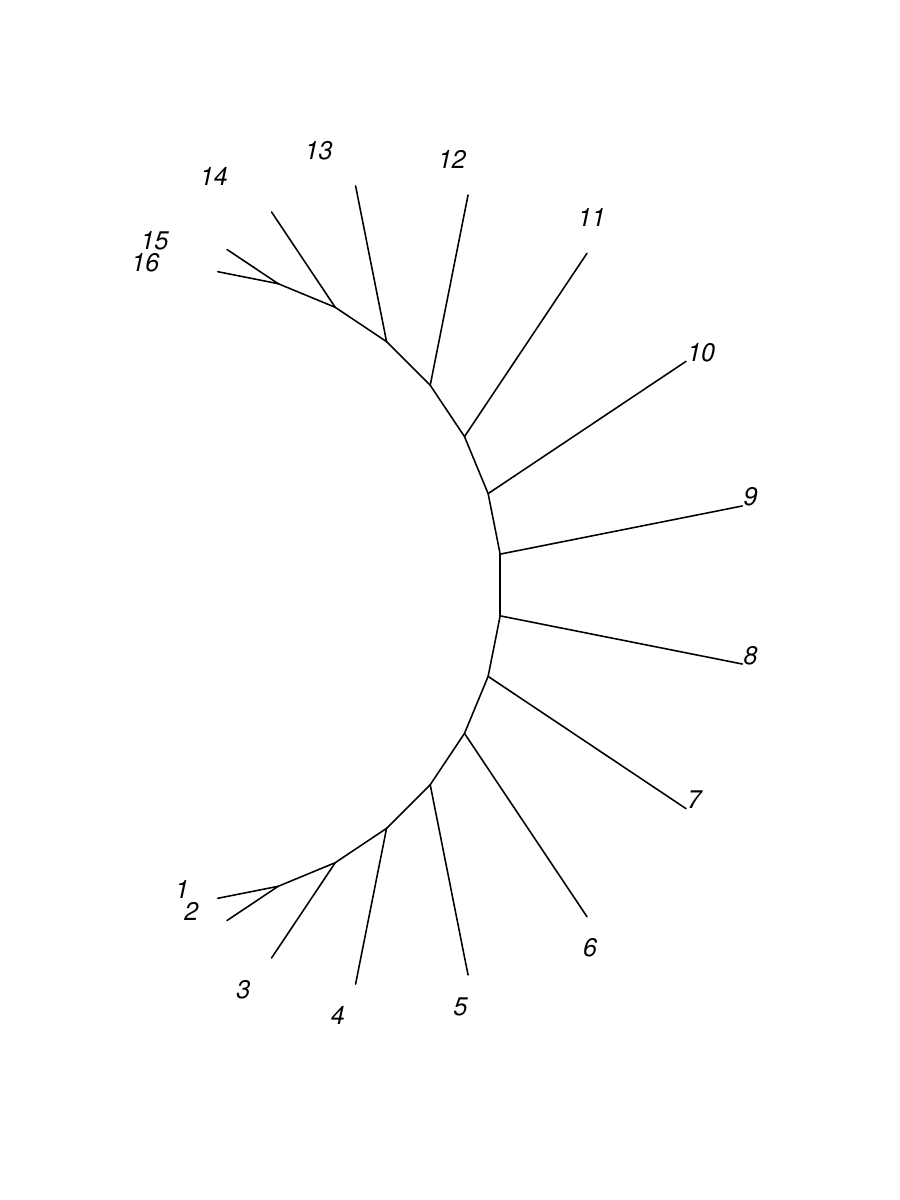}
\hskip -.3in
\includegraphics[height=2.25in]{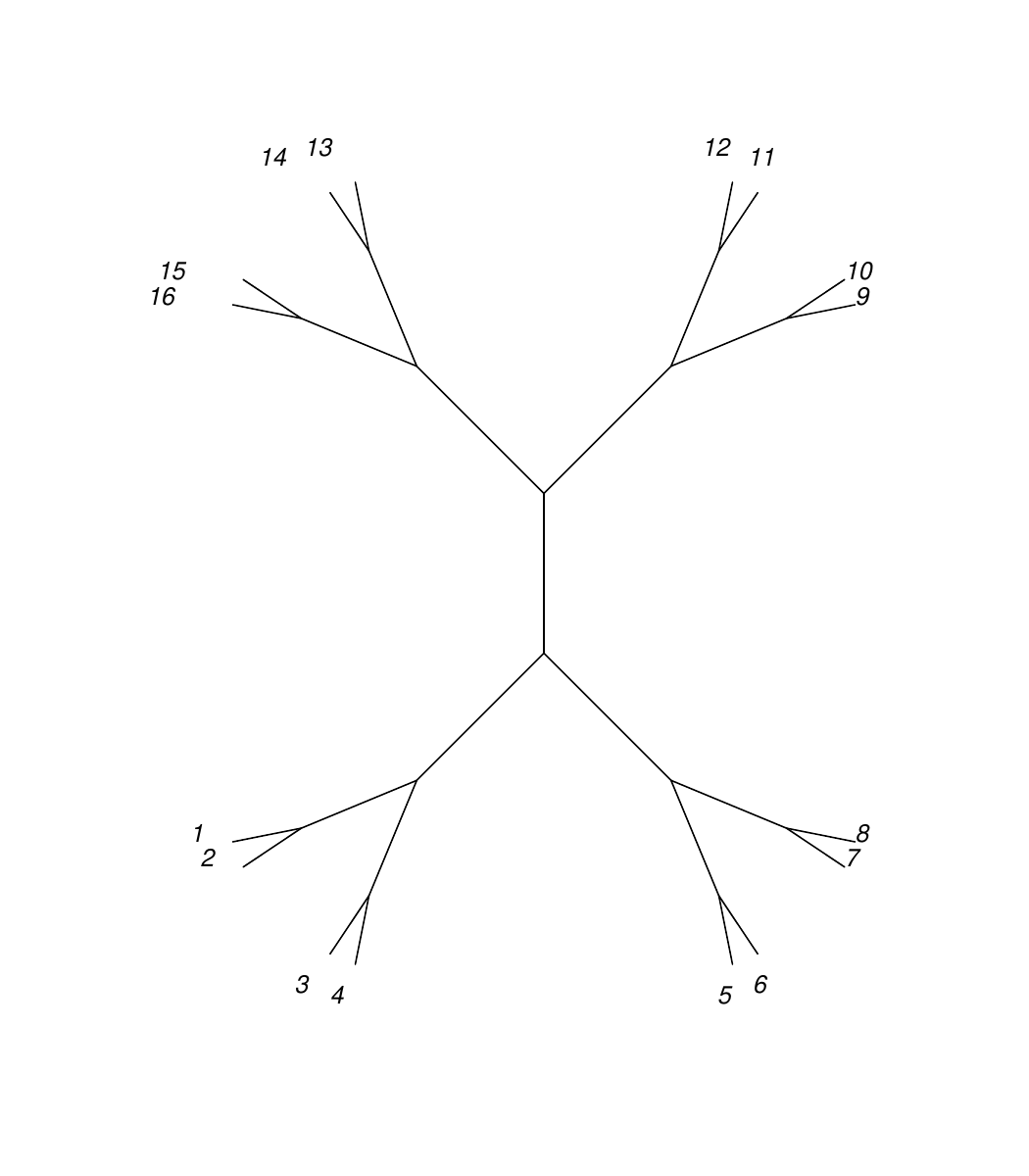}
\caption{16-taxon caterpillar and balanced trees, with edge lengths given by the quartet metrization}\label{fg:trees}
\end{figure}

\begin{example}
An unrooted caterpillar tree $$(\dots (((a_1,a_2),a_3),a_4), \dots, a_N)$$ will have internal edges inducing quartet partitions $X_1,X_2| X_3,X_4$ with sets of size
$$|X_1|=k-1,\ |X_2|=1,\ |X_3|=1,\ |X_4|=N-k-1,$$
for $k=2,3,\dots N-2$. Under the quartet metrization, the internal edge weights will thus all be $N-2$. The pendant edges to taxa $a_1,a_2,a_{N-1},a_N$, will also have weights $N-2$. Pendant edges to taxa $a_k$, for $k=3,\dots N-2$, will have weights $(k-1)(N-k)$. A 16-taxon illustration is shown in Figure \ref{fg:trees}.
\end{example}

\begin{example}
An unrooted  balanced tree 
$$(\dots ((a_1,a_2),(a_3,a_4)),\dots, ((a_{N-3},a_{N-2}),(a_{N-1},a_N))\dots )$$
on $N=2^k$ taxa will have pendant edges of weight $N-2$. With $\ell\in{1,2,\dots,k-1}$ denoting the minimal number of edges needed to connect a given internal edge to a leaf, 
the central internal edge, for which $\ell=k-1$, has weight $2^{k-2}\cdot 2^{k-2}+2^{k-2}\cdot 2^{k-2}=2^{2k-3}$, while other internal edges, with $1\le \ell\le k-2$, are of weight $$(2^{\ell-1})(2^{\ell-1})+ (2^\ell)(2^k-2^{\ell+1})=2^{k+\ell}-7\cdot 2^{2\ell-2}.$$ A 16-taxon illustration is shown in Figure \ref{fg:trees}.
\end{example}

\begin{theorem}\label{thm:dq}
Suppose an unrooted phylogenetic tree $T$ has been given the quartet metrization. Then the resulting tree metric $d_{Q}$ satisfies for all $x,y\in X$, $x\ne y$,
\begin{equation}\label{eq:dqemp}
d_{Q}(x,y)= 2|Q_{x,y}| +2N-4
\end{equation}
where $Q_{x,y}$ is the set of quartets displayed on $T$ of the form $xz|yw$ or $xyzw$.
\end{theorem}
More informally,  for a binary $T$ the distance $d_{Q}(x,y)$ is, up to a simple transformation, the number of quartets displayed on $T$ in which $x,y$ are \emph{separated}. This remains true for trees with polytomies as long as unresolved quartets are viewed as separating all their taxa.

Although Theorem \ref{thm:dq} can be deduced from Theorem \ref{thm:drt} by summing its formula over all placements of the root on pendant edges of $T$, a more direct argument is given here.
\begin{proof} Fix taxa $x\ne y$, and let $P$ denote the path in $T$ between them. 
Any node $v\ne x,y$ on $P$ determines a partition of the taxa $X=A_v\sqcup B_v\sqcup C_v^1\cdots \sqcup C_v^{k-2}$ as follows:
If $v$ has degree $k$, deleting $v$ and its incident edges partitions $X$ into $k$ non-empty subsets according to the connected components of the resulting graph. Let $A_v$ denote the partition set containing $x$, $B_v$ the one containing $y$, and  $C_v^i$ the $k-2$ remaining ones. Thus $C_v=\cup C_v^i$ contains all those taxa $z$ for which a path from $z$ to $x$ or $y$ joins $P$ at $v$.

Now any quartet $xu|yz$ or $xyuz$ that is displayed on the tree $T$ determines a node $v$ on $P$ at which the path from $u$ to $y$ joins $P$. Suppose $v$ has degree $k=k(v)$. Then the number of quartets of these forms that are displayed on $T$ and determine the same node $v$ in this way is 
\begin{multline*}
\sum_{1\le i\le k} |C_v^i|\left (\sum_{i<j\le k} |C_v^j| +|B_v|-1\right)\\
=\sum_{1\le i\le k} |C_v^i|\left (\sum_{i<j\le k} |C_v^j| +|B_v|\right) -|C_v|.
\end{multline*}
Thus 
\begin{multline}|Q_{x,y}|\\
=\sum_{v \text{ on } P\atop v\ne x,y} \left (
\sum_{1\le i\le k(v)} |C_v^i|\left (\sum_{i<j\le k(v)} |C_v^j| +|B_v|\right) -|C_v| \right ) \\
=\sum_{v \text{ on } P\atop v\ne x,y} \left (
\sum_{1\le i\le k(v)} |C_v^i|\left (\sum_{i<j\le k(v)} |C_v^j| +|B_v|\right) \right )- N+2
.\label{eq:Q1}
\end{multline}
Interchanging the roles of $x$ and $y$ we also find

\begin{multline}|Q_{x,y}|\\=
\sum_{w \text{ on } P\atop w\ne x,y} \left (
\sum_{1\le i\le k(w)} |C_w^i|\left (\sum_{i<j\le k(w)} |C_w^j| +|A_w|\right) \right )- N+2
.\label{eq:Q2}
\end{multline} 
Adding equations \eqref{eq:Q1} and \eqref{eq:Q2} and expressing the sums over $v,w$ as a single sum over edges $e=(w,v)$
in $P$ shows
\begin{equation*}2|Q_{x,y}|=\sum_{e \in P} w_Q(e)-2N+4 =d_Q(x,y)-2N+4,
\end{equation*}
which yields the claim.
\end{proof}

\begin{remark} 

Proposition 9 of \cite{BD1986} shows that  $|Q_{x,y}|+1$ for $x\ne y$ yields a tree metric on $T$, which is equivalent to the right hand side of equation \eqref{eq:dqemp} defining a tree metric on $T$. However, edge weights associated to the tree metric are not investigated in that paper. Moreover, applications of the result to tree inference, such as those discussed in Sections \ref{sec:super} and \ref{sec:st} of this work, seem not to have been pursued in intervening years.
\end{remark}

\begin{remark}
Combining Theorem \ref{thm:dq} with the fact that one can determine an unrooted metric tree from its intertaxon distances  gives an alternative proof of the well-known result that the collection of quartets displayed on a tree determines the unrooted tree topology.
\end{remark}

\begin{remark}
As a heuristic, the ASTRAL-II software \cite{ASTRALII} introduced a similarity on taxa that counts quartets \emph{not} separating two taxa on a gene tree. By Theorem \ref{thm:dq} this is essentially equivalent to the quartet metrization for that gene tree. While ASTRAL-II's goal is species tree inference,  it uses this similarity  quite differently from the quartet metrization's use in the statistically consistent approach to inference presented in Section \ref{sec:st} below.
\end{remark}

\section{Quartet Distance Supertree}\label{sec:super}

Theorems \ref{thm:drt} and \ref{thm:dq} lead to new supertree methods for finding a tree from certain collections of rooted triples or quartets. We present this fully for quartets, indicating the small modifications for rooted triples in a remark.

\smallskip

Suppose we are given a collection $\mathcal Q$ of unweighted quartets on a set of taxa $X$. We take the viewpoint that most of the given quartets show the correct phylogenetic relationship between the taxa, though some are in error. Ideally $\mathcal Q$ contains exactly one quartet for each subset of 4 taxa.

We choose a distance-based method M of tree building or selection that when applied to a tree metric on an unrooted tree $T$ returns $T$, even if $T$ is not ultrametric. We further require that its output topology is robust to small errors in the input distances at a tree metric. Possible choices for M include NJ and BioNJ (but not UPGMA) for tree building \cite{SaiNei1987,BioNJ97}, and Balanced Minimum Evolution for tree selection. In practice, a heuristic implementation known to perform well, such as FastME \cite{DespGasc2002}, may be chosen.

\begin{alg} \label{alg:QDS}(QDS/M) Quartet Distance Supertree with distance method M
 
Input: A collection $\mathcal Q$ of quartets on taxa in $ X$
\begin{enumerate}
\item For each pair $x, y\in  X$ of taxa, $x\ne y$, count the number $q(x,y)$ of quartets in $\mathcal Q$ separating $x,y$, and define the distance $\hat d_Q(x,y)=2q(x,y)+2N-4.$
\item Use the distance method M to build or select an unrooted tree from $\hat d_Q$.
\end{enumerate}
\end{alg}

\begin{remark} The name "Quartet Distance Supertree" has been chosen to emphasize its key uses of 1) displayed quartets on the input trees and 2) a distance method for constructing the supertree. Unfortunately, the term ``quartet distance" is often used to refer to a distance between two trees (based on the two trees' displayed quartets) and not the intertaxon distance (based on a collection of quartets). Since supertree methods based on finding a median tree under such an intertree distance have been explored (e.g., \cite{Ranwez2010} for a rooted-triple example), there is some potential confusion with the name chosen here. Nonetheless, the name has not been used before, and provides an accurate brief description.\end{remark}

\begin{remark} If $Q$ contains either no quartets for some sets of 4 taxa, or multiple quartets on them, 
one might view this as additional error, and modify the algorithm slightly. For instance, one might use all quartets on a given set of 4 taxa by weighting them by their relative frequency. Omitted 4-taxon subsets might be left out of counting when determining intertaxon distances, or treated as the 3 possible quartets on those taxa, each weighted by 1/3, in counting. However, these are simply hueristic adjustments. Developing any theoretical justification for them would require some model of the way in which the quartets were produced or omitted.
\end{remark}

\begin{remark}
For Rooted Triple Distance Supertree with M, one instead counts the number  $r(x,y)$ of rooted triples in a set $\mathcal R$ that separate $x,y$, and defines $\hat d_{RT}=2r(x,y)+2$. The method M can now be chosen to assume ultrametricity (e.g., UPGMA), since $\hat d_{RT}$ approximates the ultrametric tree metric $d_{RT}$. If so, then a rooted tree will be returned.
\end{remark}

Although we refer to the method of Algorithm \ref{alg:QDS} as {Quartet Distance Supertree} (QDS), and the variant for Rooted Triples as {Rooted Triple Distance Supertree} (RTDS), for a complete specification it is necessary to also indicate the distance method M used for tree construction or selection.   If none of the quartets in $\mathcal Q$ are erroneous or omitted,  QDS recovers the correct tree. However, how much error, and of what form, can occur in $\mathcal Q$ with the desired tree still accurately recovered may depend on the particular distance method M used. Since theoretical guarantees on toleration of error by distance methods tend to be much weaker than results seen in simulation studies, performance of QDS/M needs to be judged through simulation.

Assuming QDS is applied to a list of $N\choose{4}$ quartets, one for each 4-taxon subset of a $N$ taxa, the running time to produce the quartet distance matrix will be $\mathcal O(N^4)$, since considering each quartet in turn, one can increment counts for the 4 pairs of taxa that quartet separates. If Neighbor Joining, with time $\mathcal O(N^3)$, is then used to build a tree, the total complexity remains $\mathcal O(N^4)$, which is the best one can achieve for any method that
had the same input. 
While this may be too slow for some large applications, some scheme by which subsets of the quartets are sampled randomly to estimate quartet distances might still give a reasonable
approximation to the distance.

\medskip

\begin{figure}
\centering
\includegraphics[width=3in]{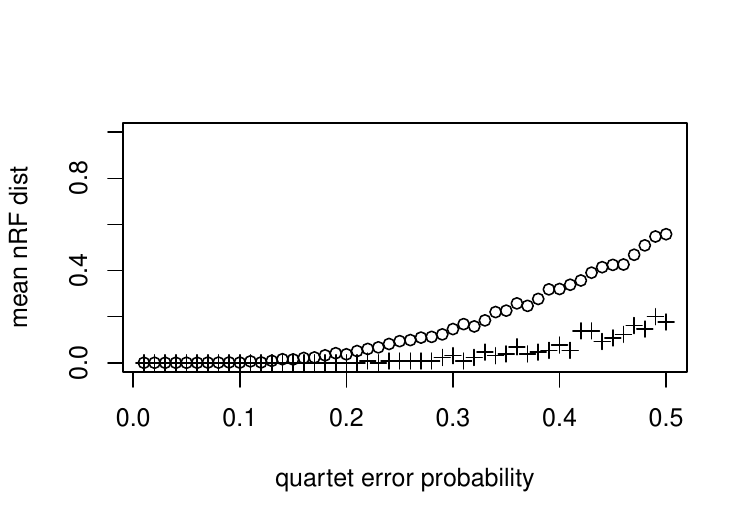}
\vskip -.1in
\includegraphics[width=3in]{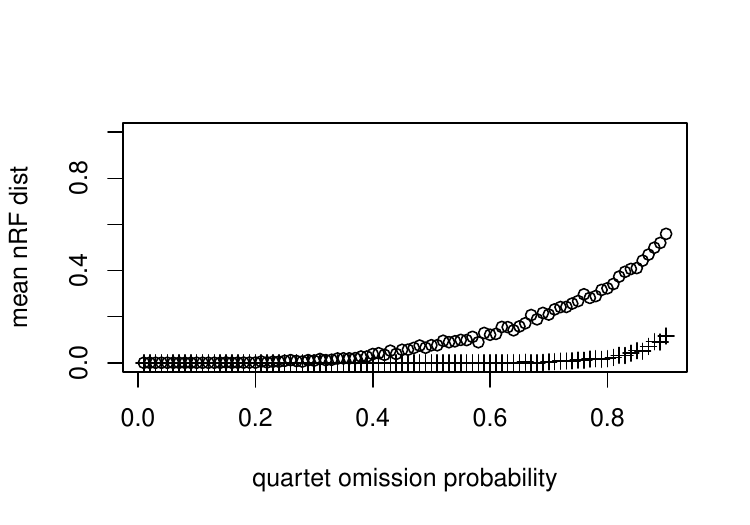}
\caption{Performance of QDS/NJ in simulation under under two scenarios, erroneous quartets (top) and omitted quartets (bottom), as described in the text. 
Circles ($\circ$) denote the 16-taxon caterpillar tree and pluses ($+$) the 16-taxon balanced tree.
The horizontal axes on the top plot gives the probability that a true quartet is replaced with an alternative on the same taxa in forming the quartet set. On the bottom plot, that axis gives the probability a quartet is omitted from the quartet set. Inference error is measured by the normalized Robinson-Foulds (nRF) distance between the correct and inferred tree.
The vertical axes show the mean nRF distance over 100 replicates.  Note that two resolved 16 taxon trees differing by one NNI  have nRF distance $2/2(16-3)\approx0.077$.}\label{fg:simul} \end{figure}

The following simulations, performed in R using the \texttt{ape} package \cite{ape2004}, give a first indiction of the performance of QDS. Using the two extreme topologies of  caterpillar and balanced trees on 16 taxa, the set of all displayed quartets was formed. Error was then introduced into the quartets in one of  two ways. In the first scenario, for choices of probability $0<p\le.5$ of quartet error, each quartet was modified with probability $p$ to one of the two resolved alternatives on the same taxa (with equal probability). In the second scenario, for choices of probability $0<p\le .9$, the quartet was removed from the set. For each of these modified quartet sets,  QDS/NJ was used to construct a tree. In the second scenario, omitted quartets were simply left out of the counting that determines intertaxon distances. The Robinson-Foulds distance was then computed between the inferred QDS/NJ tree and the original tree. This was repeated 100 times, with results summarized in the plots of Figure
\ref{fg:simul}. Similar results (not shown) were obtained using the FastME heuristic for balanced minimum evolution in place of NJ.

These results show that even with about a quarter of the quartets incorrect, on average the correct tree was recovered to within an RF distance of 2 (i.e, all but 1 of the 13 non-trivial splits were recovered correctly) for the caterpillar tree.
And even with about half of the caterpillar's quartets omitted, results were similarly accurate. The balanced tree topology was even more robustly recovered than the caterpillar tree, allowing quite large amounts of quartet error under both scenarios. 

Of course one should interpret these results cautiously, as empirical quartet error may not have the simple form of the simulation. In empirical settings it  is unlikely that all quartets would be equally likely to be incorrect or omitted, or that in the case of an incorrect quartet that both alternatives would be equally likely. 
Nonetheless, these simulations strongly indicate the need for more realistic simulation studies to investigate performance.

\begin{remark}
A potential drawback of QDS for general tree inference  from quartets is that its theoretical basis assumes one has a quartet in $\mathcal Q$ for every subset of 4 taxa, and no weights can be supplied expressing relative confidence in those quartets. This differs from the maximum quartet consistency framework in which one seeks to maximize an objective function
expressing the total weights of quartets displayed on the tree. While the above simulations suggest uniformly missing quartets may be of less concern, confidence weighting seems to be desirable, at least with quartets inferred by Maximum Likelihood,  as discussed in \cite{RanGasc2001}. However, in some applications, and especially for species tree inference from gene trees as described in the next section,
these aspects of QDS may not be a great disadvantage.
\end{remark}

\begin{remark} 
It is possible that new distance methods could be developed that are more finally tuned to QDS than those existing now. Since the distance $\hat d_Q$ approximates distances on an unknown tree $T$ endowed with the quartet metrization,  a tree building or selection method that takes that specific metrization into account may improve performance. Current distance methods are general, making no assumption about a tree's edge lengths as related to its topology.
\end{remark}

\section{Species Tree inference by Quartet Distance Consensus}\label{sec:st}

We next show how QDS and RTDS can be applied to the problem of inferring a species tree from a collection of gene trees. This provides  new consensus methods that are statistically consistent under the multispecies coalescent model, beyond those surveyed in \cite{DegnanConsensus2009}. For simplicity, we focus on the application of QDS.

For inference from multilocus sequence data,  this can be used in a two-step procedure in which  gene trees are first inferred from gene sequences, and then these inferred gene trees are treated as data for inference of a species tree. As is common for such two-stage schemes, the second stage of this method is provably statistically consistent, in the sense that if the gene trees were sampled without error under the multispecies coalescent model, then as the number of gene trees increases the probability of inferring the correct species tree approaches 1. In practice, however, there may be some inference error in the gene trees, as well as violations of the coalescent model, such as horizontal gene transfer. 

For the algorithm, we assume we already have in hand a collection of binary trees on $X$, but allow some missing taxa on each tree. However, for good performance it is desirable that each 4-taxon subset appears on many of the gene trees.

\begin{alg} (QDC/M) Quartet Distance Consensus with distance method M 

Input: A collection of  binary trees on subsets of taxa $X$
\begin{enumerate}
\item For each subset of four taxa $x,y,z,w \in X$, determine the \emph{dominant} (i.e., most frequent) quartet $xy|zw$, $xz|yw$, or $xw|yz$ displayed on the input trees. In the case of a tie, choose from the most frequent uniformly at random.
\item With $\mathcal Q$ the set of dominant quartets, apply QDS with M.
\end{enumerate}

\end{alg}

Straightforward modifications lead to a formulation of  Rooted Triple Distance Consensus (RTDC).

While QDC is similar to the clade-distance based STAR of \cite{Liu2009}, and split-distance based $\text{NJ}_{st}$ (a.k.a. U-STAR/NJ) of \cite{Liu2011}, both of those average distances across gene trees, while QDC/M instead chooses the dominant quartets across gene trees to define a distance. Note that  Rooted Triple Consensus of \cite{EwingvH2008} similarly choses the dominant rooted triple for rooted species tree inference, and inference of a population tree by the BUCKy software \cite{Larget2010} proceeds through choosing dominant quartets, though neither utilizes a distance.

\smallskip

Next we  establish statistical consistency of QDC for species tree inference under the multispecies coalescent model. This model has parameters specified by a rooted metric species tree $\sigma^r$, as described, for instance, in \cite{adr2011a}, and gives a probability distribution on binary metric gene trees. After marginalization over branch lengths and root location, one obtains a distribution on unrooted binary topological gene trees $T$. The structure of the model is such that 
one may view the generation of gene trees on subsets of taxa $Y$ as either generating gene trees under the multispecies coalescent on the induced species tree $\sigma^r_Y$ on that subset, or generating gene trees $T$ on the full species tree $\sigma^r$ and then passing to the induced gene trees $T_Y$ on the subset of taxa. We take the second approach, as it is more convenient for our argument.

By a \emph{taxon deletion model} for $X$ we mean a random variable taking as its values subsets $Y$ of $X$ . Given any tree $T$ on $X$, we apply the deletion model to $T$ by passing to the induced tree $T_Y$ on $Y$. In this formulation,  the deletion model is independent of the tree it is applied to. We call a deletion model \emph{quartet informative} if for each 4-taxon subset $F$ of $X$ the event $F\subset Y$ has positive probability.

The statistical consistency of QDC/M is then established by the following.

\begin{theorem} Let $\sigma^r$ denote a rooted binary metric species tree, with positive branch lengths, and
fix any quartet-informative taxon deletion model.
Consider a sample $S_n$ of $n$ gene trees obtained by first independently drawing gene trees on $X$ from the multispecies coalescent model on $\sigma^r$ and then applying the deletion model independently to each tree.
Let $M$ denote any tree building or selection algorithm that given pairwise distances fitting a (not necessarily ultrametric) tree will return that tree. Then with $\sigma$ the unrooted topological species tree and $\hat \sigma_n$ the unrooted topological tree inferred by QDC/M from the sample $S_n$, $$\lim_{n\to \infty} \mathbb P( \hat \sigma_n = \sigma)=1.$$
\end{theorem}

\begin{proof} Consider a subset $\{a,b,c,d\}$ of 4 taxa in $X$. Let $S_n'\subseteq S_n$ be those trees in the sample on which the 4 appear. Because the taxon deletion model is quartet informative, with probability 1 we have
$|S_n'|\to \infty$ as $n\to\infty$. Moreover, if $ab|cd$ is displayed on $\sigma^r$, the quartet on the four displayed on any tree in $S_n'$ is a trinomial random variable \cite{adr2011a} with parameters satisfying
$$p_{ab|cd}>p_{ac|bd}=p_{ad|bc}.$$
(While the precise values of these probabilities depend on branch lengths in $\sigma^r$, only the inequality and equality shown are needed
for our argument.)
Then with $$c(n)=(c_{ab|cd}(n),c_{ac|bd}(n), c_{ad|bc}(n))$$ denoting the vector of counts of the quartets displayed in $S_n'$ we have that as $n\to\infty$,
$c(n))/|S_n'|$ converges in probability to $(p_{ab|cd},p_{ac|bd},p_{ad|bc})$. This implies
$$\mathbb P( c_{ab|cd}>c_{ac|bd}, c_{ad|bc})\to 1.$$ That is, with probability approaching 1 the dominant quartet displayed on the gene trees matches that displayed on the species tree.

Since there are a finite number of subsets of 4 taxa, this implies that as $n\to\infty$ the probability approaches 1 that for \emph{all} sets of 4 taxa the dominant gene tree quartet is displayed on the species tree $\sigma$. But if all the dominant quartets are those displayed on the species tree $\sigma$, then the algorithm computes the quartet distance on $\sigma$,
so it returns $\hat \sigma_n=\sigma$.
\end{proof}

The ability to deal with missing taxa is potentially an advantage of species tree inference by QDC over the U-STAR approach. Although simulations \cite{ASTRID} have shown good performance of U-STAR with taxa missing from gene trees uniformly at random, it is unclear how relevant that pattern of missing-ness is to empirical data.The consistency of U-STAR under a uniform deletion
model is investigated in \cite{Nute2018}, but the proof given there is flawed 
(with a correction in preparation \cite{Nute2019}).
However, in the case of  non-uniform patterns of missing taxa statistical consistency seems unlikely. Indeed, it is relatively easy to construct small examples with non-uniform missing taxa where the U-STAR distance does not exactly fit any tree, though a formal proof of inconsistency would have to establish that  whatever distance method of tree construction is used cannot overcome this.

\medskip

To analyze  the running time of QDC, suppose QDC is applied to a list of $n$ gene trees, all on a set of $N$ taxa. As stated in \cite{ASTRID}, one can compute the matrix of pairwise split distances for a gene tree in time $\mathcal O(N^2)$. From this, for any 4 distinct taxa one can use the 4-point condition to determine which quartet is displayed, in constant time.  Determining all displayed quartets on a single gene tree can thus be done in time $\mathcal O(N^4)$, and counting all quartets on all $n$ gene trees in time $\mathcal O(N^4n)$. Once this is done, the dominant quartet for each set of 4 taxa separates 4 pairs of taxa, and so contributes to 4 pairwise quartet distances.
Considering each quartet, then, we obtain the pairwise distance matrix in additional time $\mathcal O(N^4)$. Using, say, NJ for tree construction requires time $\mathcal O(N^3)$, so the total time complexity of QDC/NJ is $\mathcal O(N^4n)$.

This theoretical time complexity of QDC compares poorly with $\mathcal O(N^2n+N^3)$ for $\text{NJ}_{st}$ stated in \cite{ASTRID}. The comparison to the quartet-based ASTRAL-III, however, is more interesting: In \cite{ASTRALIII}, it is stated that ASTRAL-III has time complexity $\mathcal O ((Nn)^{2.726})$ for input of binary gene trees. Thus ASTRAL-III may have better runtimes when $N>>n$ (more taxa than gene trees), but QDC may be faster for $n>>N$ (more gene trees than taxa). Since QDC has currently only been programmed in R, in a form unlikely to optimize runtime,  a better implementation of QDC is needed for a fair practical speed comparison to ASTRAL-III.

An R implementation of QDC/NJ, using the \texttt{ape} package \cite{ape2004}, on a data set of 1000 gene trees on 30 taxa took approximately 30 minutes to run on a desktop Macintosh with a 3.2GHz processor. (Of this time, over 28 minutes was spent simply tallying the displayed quartets on all the gene trees.) Although this compares poorly to approximately 22 seconds for ASTRAL-III and approximately 3 seconds for USTAR/NJ implemented in R, it still places it well within feasibility for data analysis. Indeed, the computational time to infer a large number of gene trees to serve as input will dwarf QDC's runtime. Moreover, recoding the algorithm is likely to give substantial speed improvement.

\medskip

\textbf{Simulations.} For a first look at  the possible performance of QDC for species tree inference, it was applied to Avian simulated data sets of \cite{Bayzid14}, which were also analyzed in \cite{ASTRID}.  These data sets are simulated on a fixed species tree of 48 taxa, drawn from a study of avian species. Samples of 1000 gene trees were simulated under the multispecies coalescent model on the species tree (scaling  factor 1), and on rescalings of it by $.5$ (more incomplete lineage sorting) and $2$ (less ILS). 20 replicate data sets were produced for each scaling factor. In addition to these samples of gene trees from the coalescent, sequences of length 500bp were simulated on each gene tree, and an estimated species tree inferred from them, which introduces inference error. More details on branch lengths, population sizes, and mutation rates can be found in the original publication.

\begin{figure}
\hskip -.4in
\ \ \ \ \ \ \ \ \includegraphics[width=4.2in]{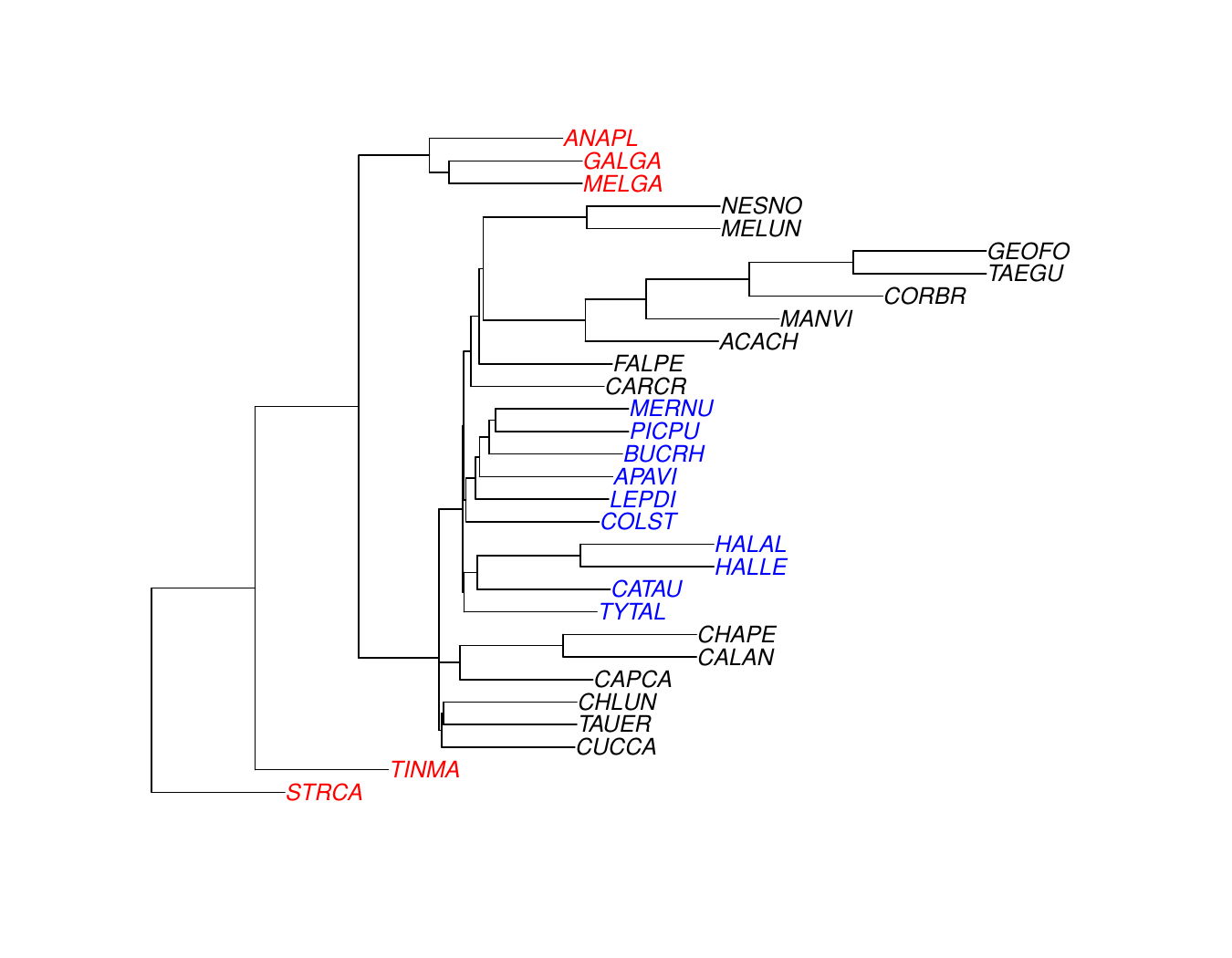}
\vskip -.6in
\caption{The 30-taxon species tree used for simulations. For trials involving missing taxa, a deletion probability $p$ was chosen, and a group of taxa deleted with probability $2p$ from each gene tree. The deleted group was the taxa shown in red (anapl, galga, melga, tinma, strca) or blue (mernu, picpu, bucrh, apavi, lepdi, colst, halal, halle, catau, tytal), with equal probability. Thus the the expected proportion of gene trees on all taxa is $1-2p$, on the the black and blue taxa is $p$, and on the black and red is $p$.}\label{fig:tree}
\end{figure}

For our simulation study, in order to reduce computational time, we reduced the number of taxa to 30, by deleting 18 taxa to obtain the species tree shown in Figure
\ref{fig:tree}. By restricting sampled gene trees on 48 taxa to the 30 chosen ones, we obtain a valid sample of gene trees from the coalescent on the restricted species tree. However, by restricting the estimated gene trees in the same way, we may not have obtained the same estimated gene tree topologies that would have been obtained from the 30 simulated sequences. Nonetheless, for a first look at performance, we expect the difference to be minor.

Data sets with missing taxa from some gene trees were also derived from these. Two subsets of the taxa, chosen as shown in Figure \ref{fig:tree} were designated.
For deletion  probabilities $p=0,$ $0.05,$ $0.1$, gene trees were chosen to have a group deleted with probability $2p$, with the particular group deleted equally likely.
Thus the expected 
proportions of gene trees missing no taxa was $1-2p$, and missing each group was $p$, with no trees missing both. 
This pattern of missing taxa was chosen to roughly mimic what might occur in empirical data sets. In particular, if the source of missing taxa on some gene trees is the biological process of gene loss, it is likely to affect closely related taxa, and if it is due to uncollected data, it might be more likely in outgroups that were not the primary focus of data collection. While our specific model of missing data is quite crude, it is perhaps more relevant than uniform-at-random deletion of individual taxa, such as was used in the simulations of \cite{ASTRID}.

As with restricting to the 30 taxa from 48, further deletion of taxa from estimated gene trees may mean they do not agree topologically with estimated gene trees from the smaller set of sequences, but differences are likely to be minor.

On these simulated data sets, we compare the performance of 5 methods based on topological features of gene trees: QDC with both balanced FastME and NJ for tree construction, U-STAR with balanced FastME and NJ, and ASTRAL-III. QDC was implemented in R (code available on request), as was U-STAR, while the more complex ASTRAL-III software was used directly.

\begin{figure*}[t!]
\centering
\includegraphics[width=5.5in]{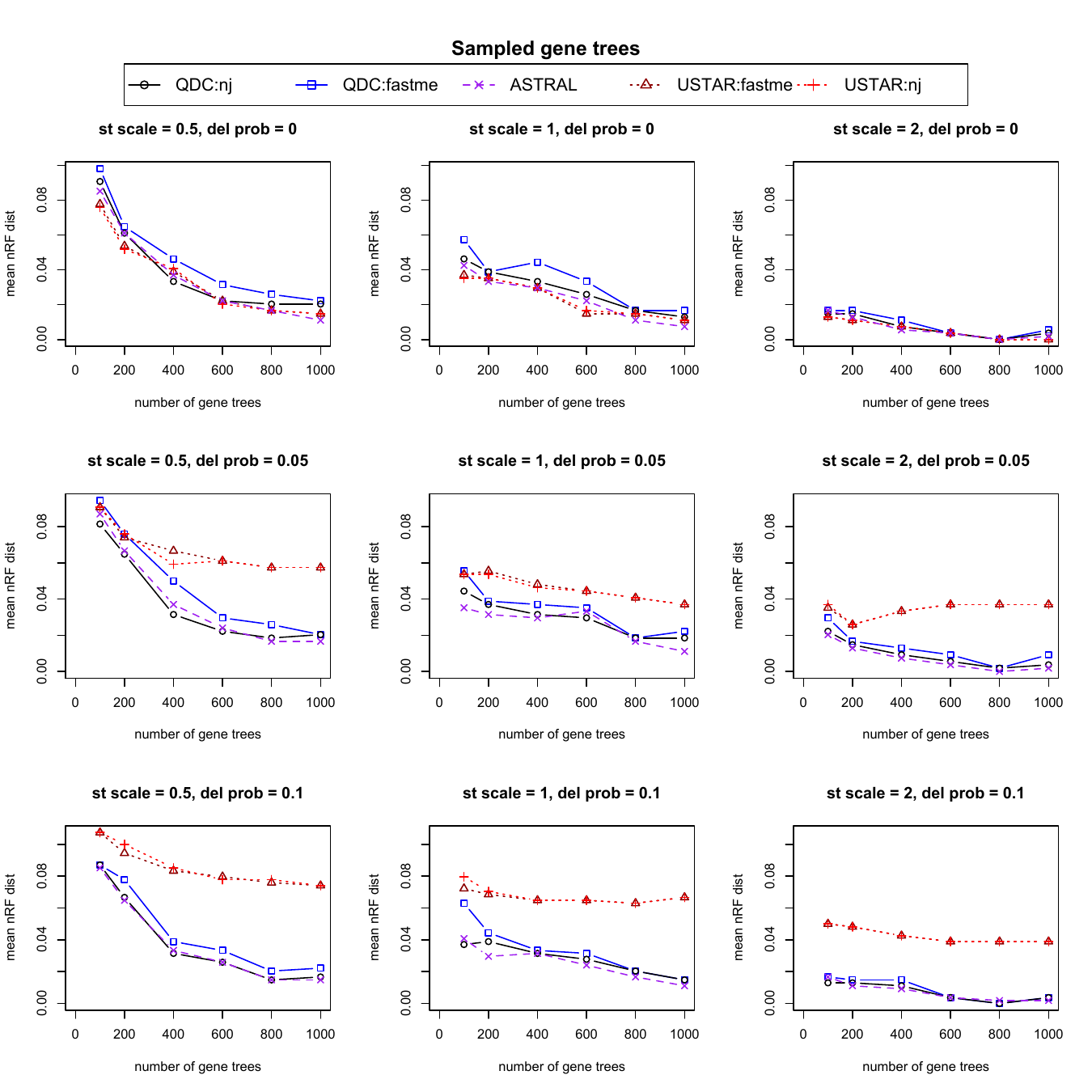}
\vskip -.1in
\caption{Simulation results based on 20 replicates for each simulation condition, using gene trees sampled from the multispecies coalescent model (lacking estimation error). From left to right, ``st scale" is the species tree scaling factor of .5,1, or 2, indicating decreasing amounts of  ILS. From top to bottom,``del prob" controls the probability of missing taxa on gene trees, with values 0, .05, or .1 indicating increasing numbers of gene trees with missing taxa. On individual plots, increasing numbers of gene trees, 100, 200, 400, 600, 800, and 1000, were analyzed for species tree inference. The mean over the replicates of the normalized Robinson-Foulds (nRF) distance from the species tree is used to measure accuracy. For 30 taxa, 2 trees differing by a single NNI move have $\text{nRF}=2/2(30-3)\approx 0.037$.
}\label{fig:true}
\end{figure*}

\begin{figure*}
\centering
\includegraphics[width=5.5in]{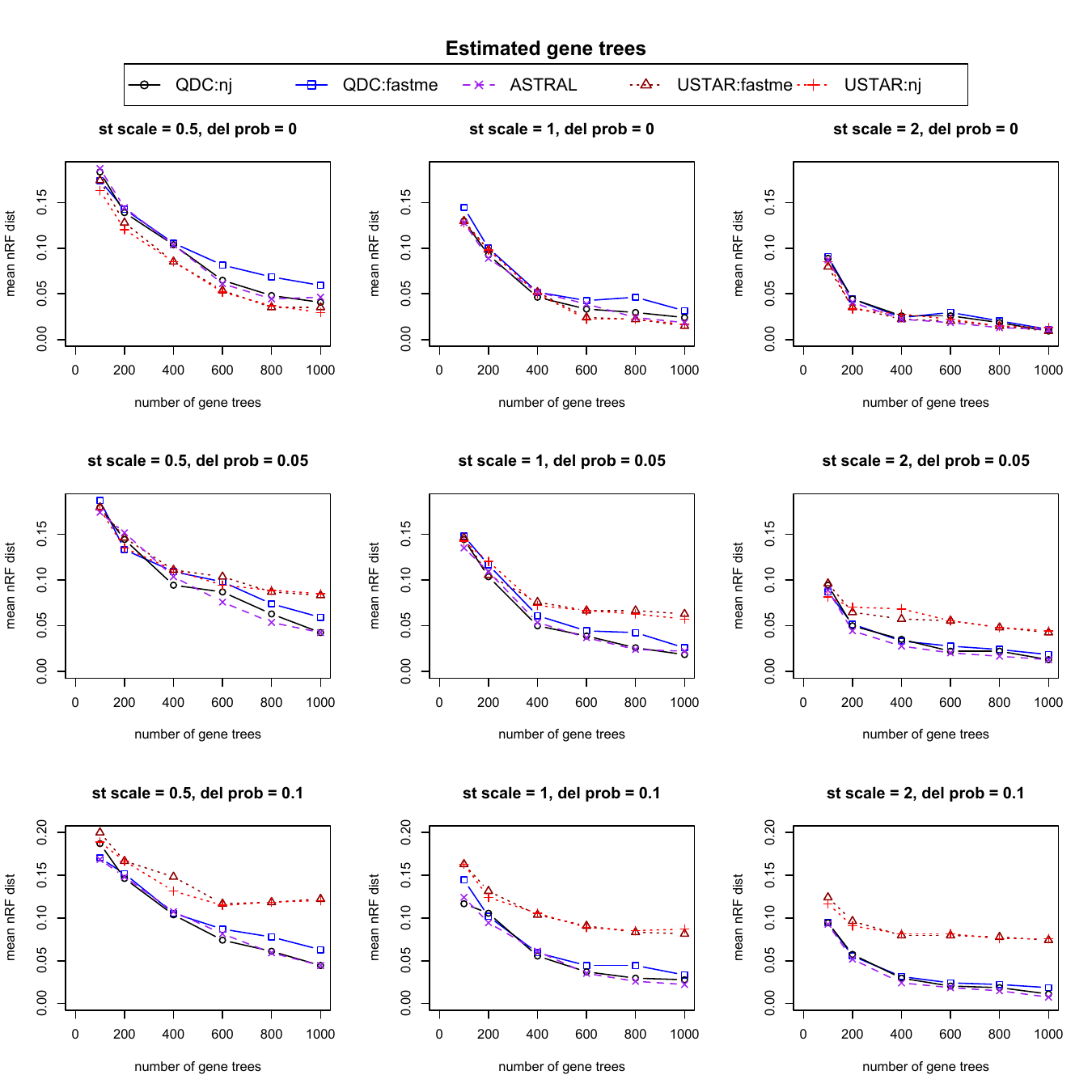}
\vskip -.1in
\caption{Simulation results based on 20 replicates for each simulation condition, using gene trees estimated from sequences simulated on gene trees sampled from the multispecies coalescent model. From left to right, ``st scale" is the species tree scaling factor of .5,1, or 2, indicating decreasing amounts of  ILS. From top to bottom,``del prob" controls the probability of missing taxa on gene trees, with values 0, .05, or .1 indicating increasing numbers of gene trees with missing taxa. On individual plots, increasing numbers of gene trees, 100, 200, 400, 600, 800, and 1000, were analyzed for species tree inference. The mean over the replicates of the normalized Robinson-Foulds (nRF) distance from the species tree is used to measure accuracy. For 30 taxa, 2 trees differing by a single NNI move have $\text{nRF}=2/2(30-3)\approx 0.037$.
}\label{fig:est}
\end{figure*}

Results are shown in Figure \ref{fig:true} using a gene tree sample under the MSC model, and in Figure \ref{fig:est} using estimated gene trees. Note that the vertical scales on all plots differ between the two figures, as species tree inference is more reliable using the sampled gene trees.

In both figures one sees that QDC performs as well or better with NJ than with FastME in all conditions. When there are no missing taxa (top row of each figure), QDC/NJ, the U-STAR methods, and ASTRAL-III perform quite similarly. Given the extra computational time QDC and ASTRAL-III require, however, these simulations show that when no taxa are missing  there is no reason to prefer QDC/NJ or ASTRAL-III to U-STAR.

When gene trees have missing taxa, however, the conclusion is quite different. At either level of missing taxa investigated (second and third rows of figures), the U-STAR methods are the poorest performing, as is in line with our earlier comments. Indeed, the plots suggest a lack of statistical consistency
of U-STAR under these conditions.
ASTRAL-III's and QDC's performance, however, do not show any pronounced changes across these missing taxon simulations, so they are clearly to be preferred to U-STAR.
Presumably, the robustness of both QDC and ASTRAL-III to missing taxa arises from their common approach of basing inference on quartets, so that if some 4-taxon sets are not on all trees, one still gets a good estimate of their relationship from the remaining trees.
Between ASTRAL-III and QDC/NJ there is little difference in performance, with no clear pattern as to which was more accurate.

While the simulations done here are by no means sufficient to draw final conclusions on the performance of QDC relative to other methods, they do indicate its potential.

\begin{remark}
The algorithm presented  and used in the simulations above implicitly assumes the species tree is binary, so that for every choice of four taxa there will be a single most probable quartet. For non-binary species trees, one may instead have all three quartets equiprobable (but not have a two-way tie for most probable). Using Theorem \ref{thm:dq} one could modify the algorithm to allow for non-binary species trees, making some choice of cutoff for judging ``near equality" in quartet frequencies.
\end{remark}

\begin{remark} 
We further note that averaging the quartet distances, or rooted triple distances, across gene trees as is done  in STAR and U-STAR would not lead to consistency under the coalescent model. In fact, an example is already given by \cite{adr2013} in which an inconsequential variant of the rooted triple metrization averaged across gene trees is shown to exactly fit an incorrect species tree for infinite sample size. A similar example for the quartet metrization is as follows:
Consider the rooted caterpillar species tree $(((((a,b)\tc x,c)\tc y,d) \tc z,e)\tc w,f)$ with $x,z,w=\infty$, $y=0$. Then under the multispecies coalescent model the gene trees $(((((a,b),c),d),e),f)$, $(((((a,b),d),c),e),f)$, and $(( ((a,b),(c,d)), e),f)$ each have probability 1/3, and all others have probability 0. Unrooting these gene trees and applying the quartet metrization, with alphabetical ordering of taxa we obtain the three distance matrices
$$\tiny \begin{pmatrix}
0&8&14&18&20&20\\
&0&14&18&20&20\\
&&0&16&18&18\\
&&&0&14&14\\
&&&&0&8\\
&&&&&0
\end{pmatrix},
\ 
\begin{pmatrix}
0&8&18&14&20&20\\
&0&18&14&20&20\\
&&0&16&14&14\\
&&&0&18&18\\
&&&&0&8\\
&&&&&0
\end{pmatrix},
$$
$$\tiny\begin{pmatrix}
0&8&18&18&18&18\\
&0&18&18&18&18\\
&&0&8&18&18\\
&&&0&18&18\\
&&&&0&8\\
&&&&&0
\end{pmatrix}.
$$
Weighting the matrices by $1/3$ and summing yields
$$\tiny\begin{pmatrix}
0&8&50/3&50/3&58/3&58/3\\
&0&50/3&50/3&58/3&58/3\\
&&0&40/3&50/3&50/3\\
&&&0&50/3&50/3\\
&&&&0&8\\
&&&&&0
\end{pmatrix},
$$
which exactly agrees with distances on the unrooted tree
$$((a\tc4,b\tc4)\tc17/3, (c\tc20/3,d\tc20/3)\tc1/3, (e\tc4,f\tc4)\tc17/3).$$ This tree does not have the same unrooted topology as the species tree. 
\end{remark}

To construct an example with a binary species tree with no zero or infinite edge lengths, we perturb the above one slightly.
If the distances on the original species tree are chosen so $x,z,w$ are very large and $y$ is very small but positive, the average of the gene tree quartet metrizations will change only slightly. Thus it cannot fit the topology of the species tree.

\section{Discussion}\label{sec:conc}

To place QDS in the context of other quartet supertree methods, note that
the most common framework in current quartet methods of inferring trees ---  maximum quartet consistency --- is to minimize an objective function measuring conflict between the given quartets and the tree.  Alternative quartet-based tree construction approaches are given in \cite{BB2001,Qjoining2008}. (Note that although our metrizations can be viewed as based in an instance of the ``isolation weighting" of \cite{BB2001}, in that work there is no concept of a true tree that one seeks to infer.)

While the optimization problem for maximum quaretet consistency should be addressed by a search over all possible trees, in practice heuristic searches are usually necessary. The number of taxa or the search space may be limited in order to achieve acceptable performance  and runtimes \cite{BaumMRP,RaganMRP,SvHQuartPuz,SnirRao10,SSLW11,ACS15,ASTRALII,ASTRALIII}.
As reasonable as this broad framework is, however, it is important to remember that the objective functions used are not ones deduced from theory. In fact, no such theory is even possible without an explicit model of
error in the quartets, and it does not appear any attempt has been made to justify current approaches in such a way. Instead, simulations which incorporate inference error in the quartets
are used for evaluation and comparison of methods.

A rather different notion of fitting a tree to quartets underlies QDS/M, whether the distance method M is a tree building algorithm or optimization of a distance-based objective function. By constructing a distance from the quartets, the selection of a ``best" tree to fit the quartets is transferred to selecting one that fits the distance.
Unfortunately no current theory can guide us as to whether this is better or worse than previous approaches.
Extensive simulation studies are needed to judge the practical effectiveness of the new methods proposed here. Moreover, since the quartet error involved in different applications may have different features, simulations studies must reflect this and be targeted at specific applications. For instance, the effectiveness of QDC for species tree inference from full $N$-taxon gene trees inferred by Maximum Likelihood (ML) may be different from that of inference by QDS of a single gene tree from quartet trees inferred by ML.

\smallskip

Quartet methods have played a role in recent progress in phylogenetics in using algebraic methods for tree inference from sequence data, in work by  \cite{SVDQsoft,SVDQtheory,CFS2007} and \cite{FSC2016}, and the ideas presented here may be useful for those applications.  Using these methods one can infer a quartet tree very quickly under very general models. However, technical issues complicate inference of larger trees directly. If QDS works well with the quartet trees these methods produce, then the significant advantages they offer, in speed and the generality of the underlying substitution model, may be broadened to include quick inference of larger trees as well.

\smallskip

For the specific problem of inference of species trees from gene trees, rooted triple and quartet approaches have been taken before by \cite{EwingvH2008,Larget2010} and \cite{ASTRALIII}, with this last work presenting the highly-developed ASTRAL-III software. While the simulations of \cite{ASTRID} suggested  ASTRAL-III's accuracy is only comparable to U-STAR, the modified analysis presented here using the same simulated data suggests that its performance is significantly more robust to missing taxa on gene trees than is U-STAR. Indeed, this feature of quartet approaches should, we believe, be more appreciated as a justification for their development. Nonetheless, in the limited simulations presented here, QDC/NJ appears to have similar performance to ASTRAL-III whether or not taxa are missing. Moreover, complexity analysis suggests that when the number of genes far exceeds the number of taxa, an efficient implementation of QDC might achieve shorter runtimes than ASTRAL-III.

\smallskip
Inference of species trees under the multispecies coalescent model is made possible by our growing ability to assemble large data sets, comprised of many genetic loci, each with its own particular genealogical history. While Bayesian methods are conceptually attractive and have been implemented to address the simultaneous inference of gene trees and species trees, with current methodology there is little hope of them giving acceptable runtimes for data sets with many taxa and loci. The QDC method proposed here takes an alternative  approach, through summarizing inferred gene trees by their displayed quartets. Compared to methods able to handle similar sized datasets, it may especially offer some gain in accuracy  in the face of missing taxa. It is based in the new QDS method of tree inference from quartets, which itself is worthy of investigation as an alternative to  methods based on the standard optimization formulation of maximum quartet consistency. While further testing of performance of these algorithms is needed, both in simulation and on empirical datasets, they offer hope for improving phylogenetic inference, and thus for helping address the many biological questions for which that is a key ingredient.

\section*{Acknowledgements}
This work was supported by the National Institutes of Health grant R01 GM117590, awarded under the  Joint DMS/NIGMS Initiative to Support Research at the Interface of the Biological and Mathematical Sciences. Thanks to Elizabeth Allman and Mike Steel, and anonymous reviewers, for comments and discussion on an earlier draft.

\bibliographystyle{plain}      
\bibliography{quartetDist}

\end{document}